\documentclass[aps,preprint]{revtex4-1}%
\usepackage{bm}
\usepackage{amsfonts}
\usepackage{amsmath}
\usepackage{amssymb}
\usepackage{graphicx}%
\begin{document}

\title{Superfluid phases of triplet pairing and neutrino emission from neutron stars}
\author{L. B. Leinson}

\begin{abstract}
Neutrino energy losses through neutral weak currents in the triplet-spin
superfluid neutron liquid are studied for the case of condensate involving
several magnetic quantum numbers. Low-energy excitations of the multicomponent
condensate in the timelike domain of the energy and momentum are analyzed.
Along with the well-known excitations in the form of broken Cooper pairs, the theoretical analysis predicts the existence of collective waves of spin density at very low energy. Because of a rather small excitation energy of spin waves, their
decay leads to a substantial neutrino emission at the lowest temperatures, when
all other mechanisms of neutrino energy loss are killed by a
superfluidity. Neutrino energy losses caused by the pair recombination and
spin-wave decays are examined in all of the multicomponent phases that might
represent the ground state of the condensate, according to modern theories, and
for the case when a phase transition occurs in the condensate at some
temperature. Our estimate predicts a sharp increase in the neutrino energy
losses followed by a decrease, along with a decrease in the temperature that takes place more rapidly
than it would without the phase transition. We demonstrate the important role of the neutrino radiation caused by the decay of spin waves in the cooling of neutron stars.

\end{abstract}
\maketitle

\affiliation{Institute of Terrestrial Magnetism, Ionosphere and Radio Wave Propagation RAS,
142190 Troitsk, Moscow Region, Russia}

\startpage{1}

\section{Introduction}

Usually neutron stars consist mostly of a superdense neutron matter which is
in $\beta$ equilibrium with a small fraction of protons and contains the
triplet-correlated superfluid condensate of neutrons below some critical
temperature \cite{Tamagaki}- \cite{Elg}. For a long time, it has been generally
accepted that the pair condensation in the superdense neutron matter occurs
into the $^{3}P_{2}$ state (with a small admixture of $^{3}F_{2}$) with a
preferred magnetic quantum number $m_{j}=0$. This model has been
conventionally used for estimates of neutrino energy losses in the minimal
cooling scenarios of neutron stars \cite{Page04}, \cite{Page09}. During the
last decade, considerable work has been done with the most realistic nuclear
potentials to determine the magnitude of the energy gap in the triplet
superfluid neutron matter for different densities \cite{Khodel}-\cite{0203046}. Sophisticated calculations have shown that, besides the above one-component
state, there are also multicomponent states involving several magnetic quantum
numbers that compete in energy and represent various phase states of the
condensate dependent on the temperature.

Whether the phase transitions modify the spectrum of low-energy excitations
and the intensity of neutrino emission from the volume of neutron stars
is the question we try to answer in this paper. Theoretical investigation of
low-energy excitations responsible for the neutrino emission by the neutron
triplet superfluid liquid is conducted first. Until recently, the only known excitations able
to decay into neutrino pairs were the broken pairs. It is
well known that the neutrino emission caused by the pair-recombination processes
in the neutron triplet superfluid liquid can dominate in the long-term
cooling of neutron stars \cite{YKL}. We will consider also the collective
excitations in the timelike domain of energies and momenta, which can also be
responsible for the intense neutrino emission. Since the neutrino emission in
the vector channel of weak interactions is strongly suppressed \cite{L10a} we
will focus on the collective spin-density oscillations that can decay into
neutrino pairs through neutral weak currents.

Previously spin modes have been studied in the $p$-wave superfluid liquid
$^{3}He$ \cite{Maki}-\cite{Wolfle}. The pairing interaction in $^{3}He$ is
invariant with respect to rotation of spin and orbital coordinates separately.
In this case, the spin fluctuations are independent of the orbital coordinates.
In contrast, the triplet-spin neutron condensate arises in high-density
neutron matter owing mostly to spin-orbit interactions that do not possess the
above symmetry. Therefore the results obtained for liquid $^{3}He$ cannot be
applied directly to the superfluid neutron liquid.

Recently spin waves with the excitation energy smaller than the superfluid
energy gap were predicted to exist in the $^{3}P_{2}$ superfluid condensate of
neutrons \cite{L10a}. The neutrino decay of such spin waves \cite{L10b} is
important for thermal evolution of neutron stars with the conventional
one-component ground state with $m_{j}=0$. In this paper, we consider
spin-density excitations for the other superfluid phases, which can be preferred at some temperatures.

We will not consider the spin oscillations of the normal component. These
soundlike waves that transfer into the ordinary spin waves in the normal
Fermi liquid above the critical temperature cannot kinematically decay into
neutrino pairs. Instead, we will focus on the spin excitations of the order
parameter, which are separated by some energy interval from the ground state
and are kinematically able to decay into neutrino pairs. The dispersion
equation for such waves in the $^{3}P_{2}$ superfluid one-component condensate
with $m_{j}=0$ was derived in Ref. \cite{L10a} in the BCS approximation. In
this paper we study the collective spin excitations in multicomponent
phases of the condensate, while taking into account the Fermi-liquid interactions.

This paper is organized as follows. Section II contains some preliminary
notes and outlines some of the important properties of the Green functions and the
one-loop integrals used below. In Sec. III we discuss the renormalization
procedure which transforms the standard gap equation to a very simple form
valid near the Fermi surface. In Sec. IV we derive the effective ordinary
and anomalous three-point vertices responsible for the interaction of the
multicomponent neutron superfluid liquid with an external axial-vector field. We
analyze the poles of anomalous vertices in order to derive the dispersion of
spin-density oscillations in the condensate. In Sec. V we derive the
linear response of the multicomponent superfluid neutron liquid onto an external axial-vector field. In Sec. VI we briefly discuss the general expression that
relates the neutrino energy losses through neutral weak currents to the
imaginary part of response functions. We derive the neutrino losses caused by
recombination of broken Cooper pairs and by decay of spin waves. Finally, in
Sec. VII, we evaluate neutrino energy losses in the multicomponent superfluid
neutron liquid undergoing the phase transition. Section VIII contains a short
summary of our findings and the conclusion.

Throughout this paper, we use the standard model of weak interactions, the system of
units $\hbar=c=1$ and the Boltzmann constant $k_{B}=1$.

\section{Preliminary notes and notation}

The spin-orbit interaction between quasiparticles is known to dominate in the
nucleon matter of high density. The most attractive channel corresponds to
spin, orbital, and total angular momenta $s=1$, $l=1$, and $j=2$, respectively, and
pairs quasiparticles into the $^{3}P_{2}$ states with $m_{j}=0,\pm1,\pm2$. The
substantially smaller tensor interactions lift the strong paramagnetic
degeneracy inherent in pure $^{3}P_{2}$ pairing and mix states of
different magnetic quantum numbers \cite{Khodel}-\cite{0203046}. The admixture
of the $^{3}F_{2}$ state, which arises because of the tensor interactions, is known
to be small and does not affect noticeably the excitation spectra \cite{L10c}.
Accordingly, throughout this paper, we neglect small tensor forces but consider
the case of pairing into the multicomponent ($m_{j}$-mixed) states
corresponding to the phases of the realistic superfluid condensate. The
pairing interaction, in the most attractive channel, can then be written as
\cite{Tamagaki} \
\begin{equation}
\varrho\Gamma_{\alpha\beta,\gamma\delta}\left(  \mathbf{p,p}^{\prime}\right)
=V\left(  p,p^{\prime}\right)  \sum_{m_{j}}\left(  \mathbf{b}_{m_{j}%
}(\mathbf{n})\hat{\bm{\sigma}}\hat{g}\right)  _{\alpha\beta}\left(  \hat
{g}\hat{\bm{\sigma}}\mathbf{b}_{m_{j}}^{\ast}(\mathbf{n}^{\prime})\right)
_{\gamma\delta}~, \label{ppint}%
\end{equation}
where $V\left(  p,p^{\prime}\right)  $ is the corresponding interaction
amplitude, $\varrho=p_{F}M^{\ast}/\pi^{2}$ is the density of states near the
Fermi surface, $\hat{\bm{\sigma}}=\left(  \hat{\sigma}_{1},\hat{\sigma}%
_{2},\hat{\sigma}_{3}\right)  $ are Pauli spin matrices, $\hat{g}=i\hat
{\sigma}_{2}$, and $\mathbf{b}_{m_{j}}\left(  \mathbf{n}\right)  $ are vectors
in spin space that generate the standard spin-angle matrices, so that
\begin{equation}
\mathbf{b}_{m_{j}}(\mathbf{n})\hat{\bm{\sigma}}\hat{g}\equiv\sum_{m_{s}%
+m_{l}=m_{j}}\left(  \frac{1}{2}\frac{1}{2}\alpha\beta|1m_{s}\right)  \left(
11m_{s}m_{l}|2m_{j}\right)  Y_{1,m_{l}}\left(  \mathbf{n}\right)  ~.
\label{bm}%
\end{equation}
These are given by
\begin{align}
\mathbf{b}_{0}  &  =\sqrt{1/2}\left(  -n_{1},-n_{2},2n_{3}\right)
~,\mathbf{b}_{1}=-\sqrt{3/4}\left(  n_{3},in_{3},n_{1}+in_{2}\right)
~,\nonumber\\
\mathbf{b}_{2}  &  =\sqrt{3/4}\left(  n_{1}+in_{2},in_{1}-n_{2},0\right)
~,\mathbf{b}_{-m_{j}}=\left(  -\right)  ^{m_{j}}\mathbf{b}_{m_{j}}^{\ast}~,
\label{b012}%
\end{align}
where $n_{1}=\sin\theta\cos\varphi$, $n_{2}=\sin\theta\sin\varphi$, and $n_{3}=\cos\theta$. The vectors are mutually orthogonal and are normalized by the
condition%
\begin{equation}
\int\frac{d\mathbf{n}}{4\pi}\mathbf{b}_{m_{j}^{\prime}}^{\ast}\mathbf{b}%
_{m_{j}}=\delta_{m_{j}m_{j}^{\prime}}. \label{lmnorm}%
\end{equation}

The triplet order parameter $\hat{D}\equiv D_{\alpha\beta}\left(
\mathbf{n}\right)  $ in the neutron superfluid represents a symmetric matrix
in spin space $\left(  \alpha,\beta=\uparrow,\downarrow\right)$, which can be
written as%
\begin{equation}
\hat{D}\left(  \mathbf{n}\right)  =\sum_{m_{j}}\Delta_{m_{j}}\left(
\hat{\bm{\sigma}}\mathbf{b}_{m_{j}}\right)  \hat{g}\ . \label{Dnlm}%
\end{equation}
We are mostly interested in the values of quasiparticle momenta $\mathbf{p}$
near the Fermi surface, $p\simeq p_{F}$, where the partial gap amplitudes
$\Delta_{m_{j}}\left(  p\right)  \simeq\Delta_{m_{j}}\left(  p_{F}\right)$
are almost constants, and the angular dependence of the order parameter is
represented by the unit vector $\mathbf{n=p}/p$, which defines the polar angles $\left(\theta,\varphi\right)$ on the Fermi surface.

The ground state (\ref{Dnlm})\ occurring in neutron matter has a relatively
simple structure (unitary triplet) \cite{Tamagaki}, \cite{Takatsuka}:
\begin{equation}
\sum_{m_{j}}\Delta_{m_{j}}\mathbf{b}_{m_{j}}\left(  \mathbf{n}\right)
=\Delta~\mathbf{\bar{b}}\left(  \mathbf{n}\right)  ~, \label{bbar}%
\end{equation}
where $\Delta$ is a complex constant (on the Fermi surface), and
$\mathbf{\bar{b}}\left(  \mathbf{n}\right)  $ is a real vector which we
normalize by the condition
\begin{equation}
\int\frac{d\mathbf{n}}{4\pi}\bar{b}^{2}\left(  \mathbf{n}\right)  =1~.
\label{Norm}%
\end{equation}
Various sets of the gap amplitudes $\Delta_{m_{j}}$ in Eq. (\ref{bbar})
correspond to the various phases of the condensate considered further.

By making use of the adopted graphical notation for the ordinary and anomalous
propagators, $\hat{G}=\parbox{1cm}{\includegraphics[width=1cm]{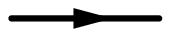}}$,
$\hat{G}^{-}(p)=\parbox{1cm}{\includegraphics[width=1cm,angle=180]{Gn.eps}}$,
$\hat{F}^{(1)}=\parbox{1cm}{\includegraphics[width=1cm]{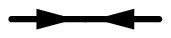}}$\thinspace,
and $\hat{F}^{(2)}=\parbox{1cm}{\includegraphics[width=1cm]{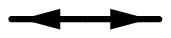}}$%
\thinspace, we employ the Matsubara calculation technique. Then the analytic
form of the propagators is as follows \cite{AGD}, \cite{Migdal}%
\begin{align}
\hat{G}\left(  \eta_{n},\mathbf{p}\right)   &  =G\left(  \eta_{n}%
,\mathbf{p}\right)  \delta_{\alpha\beta}~,\ \ \ \ \ \ \ \hat{G}^{-}\left(
\eta_{n},\mathbf{p}\right)  =G^{-}\left(  \eta_{n},\mathbf{p}\right)
\delta_{\alpha\beta}~,\nonumber\\
\hat{F}^{\left(  1\right)  }\left(  \eta_{n},\mathbf{p}\right)   &  =F\left(
\eta_{n},\mathbf{p}\right)  \mathbf{\bar{b}}\hat{\bm{\sigma}}\hat
{g}~,\ \ \ \hat{F}^{\left(  2\right)  }\left(  \eta_{n},\mathbf{p}\right)
=F\left(  \eta_{n},\mathbf{p}\right)  \hat{g}\hat{\bm{\sigma}}\mathbf{\bar{b}%
}~, \label{GF}%
\end{align}
where the scalar Green functions are of the form $G^{-}\left(  \eta
_{n},\mathbf{p}\right)  =G\left(  -\eta_{n},-\mathbf{p}\right)  $ and%
\begin{equation}
G\left(  \eta_{n},\mathbf{p}\right)  =\frac{-i\eta_{n}-\varepsilon
_{\mathbf{p}}}{\eta_{n}^{2}+E_{\mathbf{p}}^{2}}~,\ F\left(  \eta
_{n},\mathbf{p}\right)  =\frac{\Delta}{\eta_{n}^{2}+E_{\mathbf{p}}^{2}}~.
\label{GFc}%
\end{equation}
Here, $\eta_{n}\equiv i\pi\left(  2n+1\right)  T$ with $n=0,\pm1,\pm2...$ is
the fermionic Matsubara frequency and $\varepsilon_{\mathbf{p}}=\upsilon
_{F}\left(  p-p_{F}\right)  $ with $\upsilon_{F}$ the Fermi velocity.
The quasiparticle energy is given by
\begin{equation}
E_{\mathbf{p}}^{2}=\varepsilon_{\mathbf{p}}^{2}+\Delta^{2}\bar{b}^{2}\left(
\mathbf{n}\right)  ~, \label{Ep}%
\end{equation}
where the (temperature-dependent) energy gap $\Delta\bar{b}\left(
\mathbf{n}\right)  $ is anisotropic. In the absence of external fields, the
gap amplitude $\Delta\left(  T\right)  $ is real.

In general, the Green functions (\ref{GF}) should involve the renormalization
factor $a\simeq1$ independent of $\omega,\mathbf{q},T$ (see e.g.,
\cite{Migdal}). The final outcomes are independent of this factor; therefore,
to shorten the equations, we will drop the renormalization factor by assuming
that all the necessary physical values are properly renormalized.

Finally we introduce the following notation used below. We designate 
$\mathcal{I}_{XX^{\prime}}\left(  \omega,\mathbf{n,q};T\right)  $ as the
analytical continuations onto the upper-half plane of complex variable
$\omega$ of the following Matsubara sums:%
\begin{equation}
\mathcal{I}_{XX^{\prime}}\left(  \omega_{m},\mathbf{n,q};T\right)  \equiv
T\sum_{n}\frac{1}{2}\int_{-\infty}^{\infty}d\varepsilon_{\mathbf{p}}X\left(
\eta_{n}+\omega_{m},\mathbf{p+}\frac{\mathbf{q}}{2}\right)  X^{\prime}\left(
\eta_{n},\mathbf{p-}\frac{\mathbf{q}}{2}\right)  ~. \label{IXX}%
\end{equation}
where $X,X^{\prime}\in G,F,G^{-}$ and $\omega_{m}=2i\pi Tm$ with
$m=0,\pm1,\pm2,...$.These are functions of $\omega$, $\mathbf{q}$, and the
direction of the quasiparticle momentum $\mathbf{p}=p\mathbf{n}$.

The loop integrals (\ref{IXX}) possess the following properties, which can be
verified by a straightforward calculation (the same relations have been
obtained in Ref. \cite{Leggett} for the case of singlet-spin condensation):%
\begin{equation}
\mathcal{I}_{G^{-}G}=\mathcal{I}_{GG^{-}}~,~\mathcal{I}_{GF}=-\mathcal{I}%
_{FG}~,~\mathcal{I}_{G^{-}F}=-\mathcal{I}_{FG^{-}}~, \label{Leg}%
\end{equation}%
\begin{equation}
\mathcal{I}_{G^{-}F}+\mathcal{I}_{FG}=\frac{\omega}{\Delta}\mathcal{I}_{FF}~,
\label{gf}%
\end{equation}%
\begin{equation}
\mathcal{I}_{G^{-}F}-\mathcal{I}_{FG}=-\frac{\mathbf{qv}}{\Delta}%
\mathcal{I}_{FF}~. \label{ff1}%
\end{equation}
For arbitrary $\omega,\mathbf{q},T$ one can also obtain
\begin{equation}
\mathcal{I}_{GG^{-}}+\bar{b}^{2}\mathcal{I}_{FF}=A+\frac{\omega^{2}-\left(
\mathbf{qv}\right)  ^{2}}{2\Delta^{2}}\mathcal{I}_{FF}~, \label{FF}%
\end{equation}
where $\mathbf{v=}\upsilon_{F}\mathbf{n}$, and%
\begin{equation}
A\left(  \mathbf{n}\right)  \equiv\left[  \mathcal{I}_{G^{-}G}\left(
\mathbf{n}\right)  +\bar{b}^{2}\left(  \mathbf{n}\right)  \mathcal{I}%
_{FF}\left(  \mathbf{n}\right)  \right]  _{\omega=0,\mathbf{q}=0}~. \label{A}%
\end{equation}

In the case of a triplet superfluid, the key role in the response theory belongs
to the loop integrals $\mathcal{I}_{FF}$ and $\left(  \mathcal{I}_{GG}\pm
\bar{b}^{2}\mathcal{I}_{FF}\right)  $. For further usage we indicate the
properties of these functions in the case of $\omega>0$ and $\mathbf{q}
\rightarrow0 $. A straightforward calculation yields%
\begin{equation}
\lim_{q\rightarrow0}\mathcal{I}_{FF}\equiv\mathcal{I}\left(  \mathbf{n,}%
\omega\right)  =\int_{0}^{\infty}\frac{d\varepsilon}{E}\frac{\Delta^{2}%
}{4E^{2}-\left(  \omega+i0\right)  ^{2}}\tanh\frac{E}{2T}~, \label{FFq0}%
\end{equation}
and%
\begin{equation}
\left(  \mathcal{I}_{GG}+\bar{b}^{2}\mathcal{I}_{FF}\right)  _{q\rightarrow
0}=0~, \label{GGpFF}%
\end{equation}%
\begin{equation}
\left(  \mathcal{I}_{GG}-\bar{b}^{2}\mathcal{I}_{FF}\right)  _{q\rightarrow
0}=-2\bar{b}^{2}\mathcal{I}~. \label{GGmFF}%
\end{equation}

\section{Gap equation}

The standard gap equation \cite{Takatsuka} involve integration over the
regions far from the Fermi surface. This integration can be eliminated by
means of the renormalization of the pairing interaction \cite{Leggett}. We
define%
\begin{equation}
V^{\left(  r\right)  }\left(  p,p^{\prime};T\right)  =V\left(  p,p^{\prime
}\right)  -\varrho^{-1}\int\frac{dp^{\prime\prime}p^{\prime\prime2}}{\pi^{2}%
}V\left(  p,p^{\prime\prime}\right)  \left(  GG^{-}\right)  _{N}^{\prime
\prime}V^{\left(  r\right)  }\left(  p^{\prime\prime},p^{\prime};T\right)  ~,
\label{V}%
\end{equation}
where the loop $\left(  GG^{-}\right)  _{N}$ is evaluated in the normal
(nonsuperfluid) state. Then it can be shown \cite{L10a} that we may everywhere
substitute $V^{\left(  r\right)  }$ for $V$ provided that at the same time, we
understand by the $GG^{-}$ element, the subtracted quantity $GG^{-}-\left(
GG^{-}\right)  _{N}$ [$\left(  GG^{-}\right)  _{N}$ is to be evaluated for
$\omega=0,\mathbf{q}=0$ in all cases].

The function (\ref{A}) is now to be understood as%
\begin{equation}
A\left(  \mathbf{n}\right)  \rightarrow\left[  \mathcal{I}_{G^{-}%
G}-\mathcal{I}_{\left(  G^{-}G\right)  _{n}}+\bar{b}^{2}\mathcal{I}%
_{FF}\right]  _{\omega=0,\mathbf{q}=0} \label{Ar}%
\end{equation}
and the standard gap equations can be reduced to the form%
\begin{equation}
\Delta_{m_{j}}=-\Delta V^{\left(  r\right)  }\int\frac{d\mathbf{n}}{4\pi
}\mathbf{b}_{m_{j}}^{\ast}(\mathbf{n})\mathbf{\bar{b}}(\mathbf{n})A\left(
\mathbf{n}\right) ~,  \label{GAP}%
\end{equation}
which is valid in the narrow vicinity of the Fermi surface where the smooth functions
$\Delta_{m_{j}}\left(  p\right)  $, $V^{\left(  r\right)  }\left(
p,p^{\prime}\right)  $, and $\Delta\left(  p\right)  $ may be replaced with
constants $\Delta\left(  p\right)  \simeq\Delta\left(  p_{F}\right)
\equiv\Delta$, etc.

The function (\ref{Ar}) can be found explicitly after performing the
Matsubara summation:%
\begin{equation}
A\left(  \mathbf{n}\right)  =\frac{1}{2}\int_{0}^{\infty}d\varepsilon\left(
\frac{1}{\sqrt{\varepsilon^{2}+\Delta^{2}\bar{b}^{2}}}\tanh\frac
{\sqrt{\varepsilon^{2}+\Delta^{2}\bar{b}^{2}}}{2T}-\frac{1}{\varepsilon}%
\tanh\frac{\varepsilon}{2T}\right)  ~. \label{An}%
\end{equation}

\section{Effective vertices}

The field interaction with a superfluid liquid should be described with the
aid of four effective three-point vertices. There are two ordinary vertices,
$\hat{\bm{\tau}}(\mathbf{n})\mathbf{~,~}\hat{\bm{\tau}}^{-}\left(
\mathbf{n}\right)  =\hat{\bm{\tau}}^{T}(-\mathbf{n})$, corresponding to
creation of a particle and a hole by the field (which differ by the direction
of fermion lines), and two anomalous vertices, $\mathbf{\hat{T}}^{\left(
1\right)  }\left(  \mathbf{n}\right)  $ and $\mathbf{\hat{T}}^{\left(
2\right)  }\left(  \mathbf{n}\right)  $, corresponding to creation of two
particles or two holes.

\begin{figure}[h]
\includegraphics{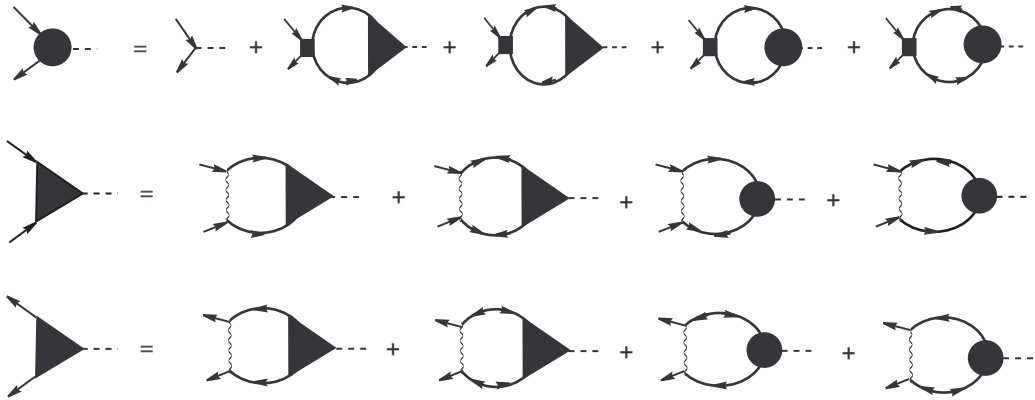}\caption{Dyson equations for full ordinary and
anomalous vertices. The particle-hole interaction is shown by the shaded
rectangle. Wavy lines represent the pairing interaction.}%
\label{fig1}%
\end{figure}

The anomalous effective vertices are given by infinite sums of the diagrams,
taking into account the pairing interaction in the ladder approximation
\cite{Nambu}. The ordinary effective vertices incorporating the particle-hole
interactions can be evaluated in the random-phase approximation \cite{Larkin}.
This can be expressed by the set of Dyson equations symbolically depicted by
graphs in Fig. \ref{fig1}.

In these diagrams the shaded circle is the full ordinary vertex,
and the shaded triangle represents the anomalous vertex. The particle-hole
interaction is shown by the shaded rectangle. Wavy lines represent the pairing
interaction. The first diagram on the right-hand side of the first line is the
three-point vertex of a free particle.

In our analysis, we shall use the fact that the Fermi-liquid interactions do
not interfere with the pairing phenomenon if approximate hole-particle
symmetry is maintained in the system, i.e., the Fermi-liquid interactions
remain unchanged upon pairing. Since we are interested in values of
quasiparticle momenta near the Fermi surface, $\mathbf{p}\simeq p_{F}%
\mathbf{n}$, the Fermi-liquid effects are reduced to the standard
particle-hole interactions:%
\[
\varrho\mathfrak{F}_{\alpha\gamma,\beta\delta}\left(  \mathbf{nn}^{\prime
}\right)  =\mathfrak{f}\left(  \mathbf{nn}^{\prime}\right)  \delta
_{\alpha\beta}\delta_{\gamma\delta}+\mathfrak{g}\left(  \mathbf{nn}^{\prime
}\right)  \bm{\sigma}_{\alpha\beta}\bm{\sigma}_{\gamma\delta}~.
\]

We are interested in excitations able to decay into neutrino pairs through
neutral weak currents. Since the neutrino emission in the vector channel of
weak interactions is strongly suppressed in nonrelativistic media \cite{L06}, \cite{L08}, \cite{L10a}, we will
focus on the interaction of the superfluid Fermi liquid with an external
axial-vector field. In the nonrelativistic case, the bare axial-vector vertex
is given by the spin matrices $\hat{\bm{\sigma}}$. (We neglect a small
temporal component that arises as the relativistic correction.)

After the proper renormalization of the pairing interaction the equations for
the axial-vector vertices can be reduced to the following analytic form (for
brevity, we omit the dependence of functions on $\omega$ and $\mathbf{q}$ ):
\begin{align}
\bm{\hat{\tau}}(\mathbf{n}) &  =\hat{\bm{\sigma}}+\hat{\bm{\sigma}}\int
\frac{d\mathbf{n}^{\prime}}{8\pi}\mathfrak{g}\left(  \mathbf{nn}^{\prime
}\right)  \left\{  \mathcal{I}_{GF}\mathrm{Tr}\left[  \hat{\bm{\sigma}}\hat
{T}^{\left(  1\right)  }\hat{g}\left(  \hat{\bm{\sigma}}\mathbf{\bar{b}%
}\right)  \right]  +\mathcal{I}_{FG}\mathrm{Tr}\left[  \hat{\bm{\sigma}}%
\left(  \hat{\bm{\sigma}}\mathbf{\bar{b}}\right)  \hat{g}\hat{T}^{\left(
2\right)  }\right]  \right.  \nonumber\\
&  \left.  +\mathcal{I}_{GG}\mathrm{Tr}\left[  \hat{\bm{\sigma}}%
\bm{\hat{\tau}}\right]  +\mathcal{I}_{FF}\mathrm{Tr}\left[  \hat
{\bm{\sigma}}\left(  \hat{\bm{\sigma}}\mathbf{\bar{b}}\right)  \hat
{g}\bm{\hat{\tau}}^{-}\hat{g}\left(  \hat{\bm{\sigma}}\mathbf{\bar{b}}\right)
\right]  \right\}  _{\mathbf{n}^{\prime}}~,\label{tau}%
\end{align}%
\begin{align}
\mathbf{\hat{T}}^{\left(  1\right)  }\left(  \mathbf{n}\right)   &
=\sum_{m_{j}}\hat{\bm{\sigma}}\mathbf{b}_{m_{j}}(\mathbf{n})\hat{g}V^{\left(
r\right)  }\int\frac{d\mathbf{n}^{\prime}}{8\pi}\left\{  \mathcal{I}_{GG^{-}%
}\mathrm{Tr}\left[  \hat{g}\left(  \hat{\bm{\sigma}}\mathbf{b}_{m_{j}}^{\ast
}\right)  \mathbf{\hat{T}}^{\left(  1\right)  }\right]  \right.  \nonumber\\
&  -\mathcal{I}_{FF}\mathrm{Tr}\left[  \left(  \hat{\bm{\sigma}}%
\mathbf{b}_{m_{j}}^{\ast}\right)  \left(  \hat{\bm{\sigma}}\mathbf{\bar{b}%
}\right)  \hat{g}\mathbf{\hat{T}}^{\left(  2\right)  }\left(  \hat
{\bm{\sigma}}\mathbf{\bar{b}}\right)  \right]  \nonumber\\
&  \left.  -\mathcal{I}_{GF}\mathrm{Tr}\left[  \left(  \hat{\bm{\sigma}}%
\mathbf{b}_{m_{j}}^{\ast}\right)  \bm{\hat{\tau}}\left(  \hat{\bm{\sigma}}%
\mathbf{\bar{b}}\right)  \right]  +\mathcal{I}_{FG^{-}}\mathrm{Tr}\left[
\left(  \hat{\bm{\sigma}}\mathbf{b}_{m_{j}}^{\ast}\right)  \left(
\hat{\bm{\sigma}}\mathbf{\bar{b}}\right)  \hat{g}\bm{\hat{\tau}}^{-}\hat
{g}\right]  \right\}  _{\mathbf{n}^{\prime}}~,\label{T1}%
\end{align}%
\begin{align}
\mathbf{\hat{T}}^{\left(  2\right)  }\left(  \mathbf{n}\right)   &
=\sum_{m_{j}}\hat{g}\hat{\bm{\sigma}}\mathbf{b}_{m_{j}}^{\ast}(\mathbf{n}%
)V^{\left(  r\right)  }\int\frac{d\mathbf{n}^{\prime}}{8\pi}\left\{
\mathcal{I}_{G^{-}G}\mathrm{Tr}\left[  \left(  \hat{\bm{\sigma}}%
\mathbf{b}_{m_{j}}\right)  \hat{g}\mathbf{\hat{T}}^{\left(  2\right)
}\right]  \right.  \nonumber\\
&  -\mathcal{I}_{FF}\mathrm{Tr}\left[  \left(  \hat{\bm{\sigma}}%
\mathbf{b}_{m_{j}}\right)  \left(  \hat{\bm{\sigma}}\mathbf{\bar{b}}\right)
\mathbf{\hat{T}}^{\left(  1\right)  }\hat{g}\left(  \hat{\bm{\sigma}}%
\mathbf{\bar{b}}\right)  \right]  \nonumber\\
&  \left.  +\mathcal{I}_{G^{-}F}\mathrm{Tr}\left[  \left(  \hat{\bm{\sigma}}%
\mathbf{b}_{m_{j}}\right)  \hat{g}\bm{\hat{\tau}}^{-}\hat{g}\left(
\hat{\bm{\sigma}}\mathbf{\bar{b}}\right)  \right]  -\mathcal{I}_{FG}%
\mathrm{Tr}\left[  \left(  \hat{\bm{\sigma}}\mathbf{b}_{m_{j}}\right)  \left(
\hat{\bm{\sigma}}\mathbf{\bar{b}}\right)  \bm{\hat{\tau}}\right]  \right\}
_{\mathbf{n}^{\prime}}~.\label{T2}%
\end{align}
Inspection of the equations reveals that the solution should be of the form%
\begin{equation}
\bm{\hat{\tau}}(\mathbf{n})=\phi\left(  \mathbf{n}\right)  \hat{\bm{\sigma}}%
\mathbf{~,~}\hat{g}\bm{\hat{\tau}}^{-}\left(  \mathbf{n}\right)  \hat{g}%
=\phi\left(  -\mathbf{n}\right)  \hat{\bm{\sigma}} \label{tauvert}%
\end{equation}%
\begin{equation}
\mathbf{\hat{T}}^{\left(  1\right)  }\left(  \mathbf{n},\omega\right)
=\sum_{m_{j}}\mathbf{B}_{m_{j}}^{\left(  1\right)  }\left(  \omega\right)
\left(  \hat{\bm{\sigma}}\mathbf{b}_{m_{j}}\right)  \hat{g}~, \label{T1A}%
\end{equation}%
\begin{equation}
\mathbf{\hat{T}}^{\left(  2\right)  }\left(  \mathbf{n},\omega\right)
=\sum_{m_{j}}\mathbf{B}_{m_{j}}^{\left(  2\right)  }\left(  \omega\right)
\hat{g}\left(  \hat{\bm{\sigma}}\mathbf{b}_{m_{j}}^{\ast}\right)  ~.
\label{T2A}%
\end{equation}
After this substitution and summation over spins, one can obtain a set of
equations for%

\begin{equation}
\phi_{\pm}\left(  \mathbf{n}\right)  =\frac{1}{2}\left(  \phi\left(
\mathbf{n}\right)  \pm\phi(-\mathbf{n})\right)  \label{fipm}%
\end{equation}
and%
\begin{equation}
\mathbf{B}_{m_{j}}^{\pm}=\frac{1}{2}\left(  \mathbf{B}_{m_{j}}^{\left(
1\right)  }\pm\left(  -\right)  ^{m_{j}}\mathbf{B}_{-m_{j}}^{\left(  2\right)
}\right)  ~. \label{Bpm}%
\end{equation}
The application of a little algebra using Eqs. (\ref{Leg})--(\ref{ff1}) results in the following equations:
\begin{align}
\phi_{+}\left(  \mathbf{n}\right)   &  =1+\int\frac{d\mathbf{n}^{\prime}}%
{4\pi}\mathfrak{g}\left(  \mathbf{nn}^{\prime}\right)  \left(  \mathcal{I}%
_{GG}-\bar{b}^{2}\mathcal{I}_{FF}\right)  _{\mathbf{n}^{\prime}}\phi
_{+}\left(  \mathbf{n}^{\prime}\right) \nonumber\\
&  -\sum_{m_{j}}\mathbf{B}_{m_{j}}^{-}\frac{\omega}{\Delta}\int\frac
{d\mathbf{n}^{\prime}}{4\pi}\mathfrak{g}\left(  \mathbf{nn}^{\prime}\right)
\mathcal{I}_{FF}\left(  \mathbf{n}^{\prime}\right)  i\left(  \mathbf{b}%
_{m_{j}}\mathbf{\times\bar{b}}\right)  _{\mathbf{n}^{\prime}}~, \label{fip}%
\end{align}%
\begin{align}
\phi_{-}\left(  \mathbf{n}\right)   &  =\int\frac{d\mathbf{n}^{\prime}}{4\pi
}\mathfrak{g}\left(  \mathbf{nn}^{\prime}\right)  \left(  \mathcal{I}%
_{GG}+\bar{b}^{2}\mathcal{I}_{FF}\right)  _{\mathbf{n}^{\prime}}\phi
_{-}\left(  \mathbf{n}^{\prime}\right) \nonumber\\
&  +\sum_{m_{j}}\mathbf{B}_{m_{j}}^{+}\int\frac{d\mathbf{n}^{\prime}}{4\pi
}\mathfrak{g}\left(  \mathbf{nn}^{\prime}\right)  \frac{\mathbf{qv}}{\Delta
}\mathcal{I}_{FF}\left(  \mathbf{n}^{\prime}\right)  i\left(  \mathbf{b}%
_{m_{j}}\mathbf{\times\bar{b}}\right)  _{\mathbf{n}^{\prime}}~, \label{fim}%
\end{align}%
\begin{align}
&  \sum_{m_{j}^{\prime}}\mathbf{B}_{m_{j}^{\prime}}^{+}\left[  \int
\frac{d\mathbf{n}}{4\pi}\left(  \mathbf{b}_{m_{j}}^{\ast}\mathbf{b}%
_{m_{j}^{\prime}}-\delta_{m_{j},m_{j}^{\prime}}\frac{\Delta}{\Delta_{m_{j}}%
}\mathbf{b}_{m_{j}}^{\ast}\mathbf{\bar{b}}\right)  A\right. \nonumber\\
&  \left.  +\int\frac{d\mathbf{n}}{4\pi}\left(  \frac{\omega^{2}-\left(
\mathbf{qv}\right)  ^{2}}{2\Delta^{2}}\mathbf{b}_{m_{j}}^{\ast}\mathbf{b}%
_{m_{j}^{\prime}}-2\left(  \mathbf{b}_{m_{j}}^{\ast}\mathbf{\bar{b}}\right)
\left(  \mathbf{\bar{b}b}_{m_{j}^{\prime}}\right)  \right)  \mathcal{I}%
_{FF}\right] \nonumber\\
&  =-2\int\frac{d\mathbf{n}}{4\pi}\left(  \frac{\omega}{\Delta}\phi_{+}%
+\frac{\mathbf{qv}}{\Delta}\phi_{-}\right)  i\left(  \mathbf{b}_{m_{j}}^{\ast
}\mathbf{\times\bar{b}}\right)  \mathcal{I}_{FF}~, \label{Bp}%
\end{align}%
\begin{gather}
\sum_{m_{j}^{\prime}}\mathbf{B}_{m_{j}^{\prime}}^{-}\left[  \int
\frac{d\mathbf{n}}{4\pi}\left(  \mathbf{b}_{m_{j}}^{\ast}\mathbf{b}%
_{m_{j}^{\prime}}-\delta_{m_{j}m_{j}^{\prime}}\frac{\Delta}{\Delta_{m_{j}}%
}\mathbf{b}_{m_{j}}^{\ast}\mathbf{\bar{b}}\right)  A\right. \nonumber\\
\left.  +\int\frac{d\mathbf{n}}{4\pi}\left(  \left(  \frac{\omega^{2}-\left(
\mathbf{qv}\right)  ^{2}}{2\Delta^{2}}-2\bar{b}^{2}\right)  \mathbf{b}_{m_{j}%
}^{\ast}\mathbf{b}_{m_{j}^{\prime}}+2\left(  \mathbf{b}_{m_{j}}^{\ast
}\mathbf{\bar{b}}\right)  \left(  \mathbf{\bar{b}b}_{m_{j}^{\prime}}\right)
\right)  \mathcal{I}_{FF}\right]  =0~. \label{Bm}%
\end{gather}
In obtaining the last two equations, we used the gap equation (\ref{GAP}) and
the identity (\ref{FF}).

Further simplifications are possible due to the fact that $\mathbf{B}_{m_{j}%
}^{-}$ in Eq. (\ref{Bm}) do not couple to external fields. Even if the eigenoscillations of $\mathbf{B}_{m_{j}}^{-}$ exist it is unclear how this mode
could be excited. Therefore one may assume that Eq. (\ref{Bm}) has only
the trivial solution $\mathbf{B}_{m_{j}}^{-}=0$. This simplifies Eq. (\ref{fip})
which is now uncoupled.

The amplitudes of Fermi-liquid interactions can be expanded into Legendre
polynomials and written in terms of an infinite set of Landau parameters. In the axial channel, this gives
\begin{equation}
\mathfrak{g}\left(  \mathbf{nn}^{\prime}\right)  =\sum_{l=0}^{\infty
}\mathfrak{g}_{l}P_{l}\left(  \mathbf{nn}^{\prime}\right)  ~. \label{lexp}%
\end{equation}
We now expand the functions $\phi_{\pm}\left(  \mathbf{n}\right)  $ over
spherical harmonics $Y_{lm}\left(  \mathbf{n}\right)  $. It is apparent that the
function $\phi_{+}\left(  \mathbf{n}\right)  =\phi_{+}\left(  -\mathbf{n}%
\right)  $ contains only even harmonics,%
\begin{equation}
\phi_{+}\left(  \mathbf{n};\omega,\mathbf{q}\right)  =\sqrt{4\pi}%
\sum_{l=\mathsf{even}}\sum_{m=-l}^{l}\phi_{l,m}^{+}\left(  \omega
,\mathbf{q}\right)  Y_{l,m}\left(  \mathbf{n}\right)  ~, \label{fipY}%
\end{equation}
while $\phi_{-}\left(  \mathbf{n}\right)  =-\phi_{-}\left(  -\mathbf{n}%
\right)  $ consists of odd harmonics,
\begin{equation}
\phi_{-}\left(  \mathbf{n};\omega,\mathbf{q}\right)  =\sqrt{4\pi}%
\sum_{l=\mathsf{odd}}\sum_{m=-l}^{l}\phi_{l,m}^{-}\left(  \omega
,\mathbf{q}\right)  Y_{l,m}\left(  \mathbf{n}\right)  ~. \label{fimY}%
\end{equation}
Making use of the relation%
\[
\int d\mathbf{n}Y_{\lambda,\mu}^{\ast}\left(  \mathbf{n}\right) \mathfrak{g}\left(
\mathbf{nn}^{\prime}\right)  =\sum_{l}\frac{4\pi}{2l+1}\mathfrak{g}_{l}\delta_{\lambda
,l}Y_{l,\mu}^{\ast}\left(  \mathbf{n}^{\prime}\right)  ~,
\]
which follows from the expansion (\ref{lexp}), we arrive at the final set of
of equations,
\begin{equation}
\phi_{l,m}^{+}=\delta_{l,0}\delta_{m,0}+\frac{\mathfrak{g}_{l}}{2l+1}%
\sum_{l^{\prime}=\mathsf{even}}\sum_{m^{\prime}}\phi_{l^{\prime},m^{\prime}%
}^{+}\int d\mathbf{n}Y_{l,m}^{\ast}\left(  \mathcal{I}_{GG}-\bar{b}%
^{2}\mathcal{I}_{FF}\right)  Y_{l^{\prime},m^{\prime}}~, \label{fipEq}%
\end{equation}%
\begin{align}
\phi_{l,m}^{-}  &  =\frac{\mathfrak{g}_{l}}{2l+1}\sum_{l^{\prime}%
=\mathsf{odd}}\sum_{m^{\prime}}\phi_{l^{\prime},m^{\prime}}^{-}\int
d\mathbf{n}Y_{l,m}^{\ast}\left(  \mathcal{I}_{GG}+\bar{b}^{2}\mathcal{I}%
_{FF}\right)  Y_{l^{\prime},m^{\prime}}\nonumber\\
&  +\frac{\sqrt{4\pi}\mathfrak{g}_{l}}{2l+1}i\frac{q\upsilon_{F}}{\Delta}%
\sum_{m_{j}}\mathbf{B}_{m_{j}}^{+}\int\frac{d\mathbf{n}}{4\pi}\left(
\cos\theta_{\mathbf{qn}}\right)  \left(  \mathbf{b}_{m_{j}}\mathbf{\times
\bar{b}}\right)  \mathcal{I}_{FF}Y_{l,m}^{\ast}~, \label{fimEq}%
\end{align}%
\begin{align}
&  \sum_{m_{j}^{\prime}}\mathbf{B}_{m_{j}^{\prime}}^{+}\left[  \int
\frac{d\mathbf{n}}{4\pi}\left(  \mathbf{b}_{m_{j}}^{\ast}\mathbf{b}%
_{m_{j}^{\prime}}-\delta_{m_{j},m_{j}^{\prime}}\frac{\Delta}{\Delta_{m_{j}}%
}\mathbf{b}_{m_{j}}^{\ast}\mathbf{\bar{b}}\right)  A\right. \nonumber\\
&  \left.  +\int\frac{d\mathbf{n}}{4\pi}\left(  \frac{\omega^{2}-q^{2}%
\upsilon_{F}^{2}\left(  \cos\theta_{\mathbf{qn}}\right)  ^{2}}{2\Delta^{2}%
}\mathbf{b}_{m_{j}}^{\ast}\mathbf{b}_{m_{j}^{\prime}}-2\left(  \mathbf{b}%
_{m_{j}}^{\ast}\mathbf{\bar{b}}\right)  \left(  \mathbf{\bar{b}b}%
_{m_{j}^{\prime}}\right)  \right)  \mathcal{I}_{FF}\right] \nonumber\\
&  =-2i\sqrt{4\pi}\sum_{l=\mathsf{even}}\sum_{m}\phi_{l,m}^{+}\frac{\omega
}{\Delta}\int\frac{d\mathbf{n}}{4\pi}\left(  \mathbf{b}_{m_{j}}^{\ast
}\mathbf{\times\bar{b}}\right)  \mathcal{I}_{FF}Y_{l,m}\nonumber\\
&  -2i\sqrt{4\pi}\sum_{l=\mathsf{odd}}\sum_{m}\phi_{l,m}^{-}\frac
{q\upsilon_{F}}{\Delta}\int\frac{d\mathbf{n}}{4\pi}\left(  \cos\theta
_{\mathbf{qn}}\right)  \left(  \mathbf{b}_{m_{j}}^{\ast}\mathbf{\times\bar{b}%
}\right)  \mathcal{I}_{FF}Y_{l,m}~, \label{BpEq}%
\end{align}
where $\theta_{\mathbf{qn}}$ is the angle between the transferred momentum and
the direction of quasiparticle motion.

Since $l$ can take all values from zero to infinity, a general solution cannot
be given in closed form. As in the case of a normal Fermi-liquid, a closed
solution may be obtained if we set $\mathfrak{g}_{l}=0$ for $l>1$. We adopt
this approximation and consider the solutions with $\mathbf{q=0}$.

With the aid of Eqs. (\ref{FFq0})--(\ref{GGmFF}), we find $\phi_{l,m}^{-}=0$
and
\begin{equation}
\phi_{0,0}^{+}=\frac{1}{1+2\mathfrak{g}_{0}\left\langle \bar{b}^{2}%
\mathcal{I}\left(  \mathbf{n};\omega\right)  \right\rangle }~;~\phi_{l,m}%
^{+}=0~,~l>0 ~. \label{fip00}%
\end{equation}
Hereafter the angle brackets denote angle averages, $\left\langle
...\right\rangle \equiv\left(  4\pi\right)  ^{-1}\int d\mathbf{n}...$.
Inserting these functions in Eq. (\ref{BpEq}) we obtain the set of equations
for $\mathbf{B}_{m_{j}}^{+}$,%
\begin{align}
&  \sum_{m_{j}^{\prime}}\mathbf{B}_{m_{j}^{\prime}}^{+}\left[  \left\langle
\left(  \mathbf{b}_{m_{j}}^{\ast}\mathbf{b}_{m_{j}^{\prime}}\right)
A\right\rangle -\delta_{m_{j},m_{j}^{\prime}}\frac{\Delta}{\Delta_{m_{j}}%
}\left\langle \left(  \mathbf{b}_{m_{j}}^{\ast}\mathbf{\bar{b}}\right)
A\right\rangle \right. \nonumber\\
&  \left.  +\frac{\omega^{2}}{2\Delta^{2}}\left\langle \left(  \mathbf{b}%
_{m_{j}}^{\ast}\mathbf{b}_{m_{j}^{\prime}}\right)  \mathcal{I}\right\rangle
-2\left\langle \left(  \mathbf{b}_{m_{j}}^{\ast}\mathbf{\bar{b}}\right)
\left(  \mathbf{\bar{b}b}_{m_{j}^{\prime}}\right)  \mathcal{I}\right\rangle
\right] \nonumber\\
&  =\frac{-2i}{1+2\mathfrak{g}_{0}\left\langle \bar{b}^{2}\mathcal{I}%
\right\rangle }\frac{\omega}{\Delta}\left\langle \left(  \mathbf{b}_{m_{j}%
}^{\ast}\mathbf{\times\bar{b}}\right)  \mathcal{I}\right\rangle ~.
\label{Bplus}%
\end{align}
The unit vector $\mathbf{\bar{b}}$ used here was defined by Eq.
(\ref{bbar}).

The explicit evaluation of Eq. (\ref{Bplus}) for arbitrary values of $\omega$ and
$T$ appears to require numerical computation. However, we can get a clear idea
of the behavior of this function using the angle-averaged energy gap
$\Delta^{2}\bar{b}^{2}\rightarrow\left\langle \Delta^{2}\bar{b}^{2}%
\right\rangle =$ $\Delta^{2}$ in the quasiparticle energy (\ref{Ep}).
(Replacing the angle-dependent gap in the quasiparticle energy by its average has
been found to be a good approximation \cite{Baldo}, \cite{L10a}, \cite{L10c}.)
In this approximation, the functions $\mathcal{I}\left(\omega,T;
\mathbf{n}\right)
\rightarrow\mathcal{I}_{\mathsf{av}}\left(  \omega,T\right)$ and $A\left(
T\right)  $, in Eqs. (\ref{fip00}) and (\ref{Bplus}), can be moved beyond the
angle integrals. With the aid of normalization condition (\ref{Norm}), we find%
\begin{equation}
\phi_{0,0}^{+}\left(  \omega,T\right)  =\frac{1}{1+2\mathfrak{g}%
_{0}\mathcal{I}_{\mathsf{av}}}~.\label{fi00p}%
\end{equation}
Using also the fact that
\begin{equation}
\left\langle \mathbf{b}_{m_{j}}^{\ast}\mathbf{b}_{m_{j}^{\prime}}\right\rangle
=\delta_{m_{j}m_{j}^{\prime}}~,~\frac{\Delta}{\Delta_{m_{j}}}\left\langle
\mathbf{b}_{m_{j}}^{\ast}\mathbf{\bar{b}}\right\rangle =1\label{Am}%
\end{equation}
and substituting%
\[
\mathbf{B}_{m_{j}}^{+}\equiv\frac{1}{1+2\mathfrak{g}_{0}\mathcal{I}%
_{\mathsf{av}}}\mathbf{B}_{m_{j}}~,
\]
from Eq. (\ref{Bplus}), we obtain the set of linear equations for
$\mathbf{B}_{m_{j}}$ with $m_{j}=0,\pm1,\pm2$,
\begin{equation}
\frac{\omega^{2}}{4\Delta^{2}}\mathbf{B}_{m_{j}}-\sum_{m_{j}^{\prime}%
}\mathbf{B}_{m_{j}^{\prime}}\left\langle \left(  \mathbf{b}_{m_{j}}^{\ast
}\mathbf{\bar{b}}\right)  \left(  \mathbf{\bar{b}b}_{m_{j}^{\prime}}\right)
\right\rangle =-i\frac{\omega}{\Delta}\left\langle \mathbf{b}_{m_{j}}^{\ast
}\mathbf{\times\bar{b}}\right\rangle ~.\label{Beq}%
\end{equation}
For further progress, we need to define a particular form of vector
$\mathbf{\bar{b}}$ that characterizes the ground state of the condensate. The
general form of a unitary $^{3}P_{2}$ state is to be written as
\begin{equation}
\mathbf{\bar{b}}=\frac{\Delta_{0}}{\Delta}\mathbf{b}_{0}+\frac{\Delta_{1}%
}{\Delta}\left(  \mathbf{b}_{1}-\mathbf{b}_{-1}\right)  +\frac{\Delta_{2}%
}{\Delta}\left(  \mathbf{b}_{2}+\mathbf{b}_{-2}\right)  \label{unitary}%
\end{equation}
with
\begin{equation}
\Delta^{2}=\Delta_{0}^{2}+2\Delta_{1}^{2}+2\Delta_{2}^{2}~.\label{del}%
\end{equation}
By utilizing notation adopted in Refs. \cite{Khodel}, \cite{Clark}, where
$\lambda_{1}\equiv\sqrt{6}\Delta_{1}/\Delta_{0}$ and $\lambda_{2}\equiv
\sqrt{6}\Delta_{2}/\Delta_{0}$, from Eq. (\ref{unitary}) we obtain the general
form of the properly normalized vector $\mathbf{\bar{b}}$:
\begin{equation}
\mathbf{\bar{b}}=\sqrt{\frac{1}{2}}\frac{\Delta_{0}}{\Delta}\left(
\begin{array}
[c]{ccc}%
-n_{1}+n_{1}\lambda_{2}-n_{3}\lambda_{1}~, & -n_{2}-n_{2}\lambda_{2}~, &
2n_{3}-n_{1}\lambda_{1}%
\end{array}
\right)  ~.\label{burb}%
\end{equation}
The solution to the set of linear equations (\ref{Beq}) is
found to be
\begin{equation}
\mathbf{B}_{0}=-\frac{1}{2}\bar{\omega}\frac{\Delta_{0}^{3}}{\Delta^{3}}%
\frac{\lambda_{1}}{\bar{\omega}^{2}-\bar{\omega}_{0}^{2}}\left(  0,i,0\right)
\label{B0}%
\end{equation}%
\begin{equation}
\mathbf{B}_{\pm1}=\frac{1}{4}\sqrt{\frac{2}{3}}\bar{\omega}\frac{\Delta
_{0}^{3}}{\Delta^{3}}\left(  -\frac{\left(  3+\lambda_{2}\right)  \left(
\bar{\omega}^{2}-\bar{\omega}_{3}^{2}\right)  }{\left(  \bar{\omega}^{2}%
-\bar{\omega}_{1}^{2}\right)  \left(  \bar{\omega}^{2}-\bar{\omega}_{2}%
^{2}\right)  },\frac{\pm i\left(  3-\lambda_{2}\right)  }{\bar{\omega}%
^{2}-\bar{\omega}_{0}^{2}},-\frac{\lambda_{1}\left(  \bar{\omega}^{2}%
-\bar{\omega}_{4}^{2}\right)  }{\left(  \bar{\omega}^{2}-\bar{\omega}_{1}%
^{2}\right)  \left(  \bar{\omega}^{2}-\bar{\omega}_{2}^{2}\right)  }\right)
\label{B1}%
\end{equation}%
\begin{equation}
\mathbf{B}_{\pm2}=\mp\frac{1}{4}\sqrt{\frac{2}{3}}\bar{\omega}\frac{\Delta
_{0}^{3}}{\Delta^{3}}\left(  \frac{\lambda_{1}\left(  \bar{\omega}^{2}%
-\bar{\omega}_{4}^{2}\right)  }{\left(  \bar{\omega}^{2}-\bar{\omega}_{1}%
^{2}\right)  \left(  \bar{\omega}^{2}-\bar{\omega}_{2}^{2}\right)  },\frac{\pm
i\lambda_{1}}{\bar{\omega}^{2}-\bar{\omega}_{0}^{2}},\frac{2\lambda_{2}\left(
\bar{\omega}^{2}-\bar{\omega}_{5}^{2}\right)  }{\left(  \bar{\omega}^{2}%
-\bar{\omega}_{1}^{2}\right)  \left(  \bar{\omega}^{2}-\bar{\omega}_{2}%
^{2}\right)  }\right)  ~,\label{B2}%
\end{equation}
where $\bar{\omega}\equiv\omega/\left(  2\Delta\right)$, and we use the
following notation:
\begin{equation}
\bar{\omega}_{0}^{2}=\frac{1}{20}\frac{\Delta_{0}^{2}}{\Delta^{2}}\left(
1+\lambda_{2}\right)  ^{2}~,\label{w1}%
\end{equation}%
\begin{equation}
\bar{\omega}_{1}^{2}=\frac{1}{40}\frac{\Delta_{0}^{2}}{\Delta^{2}}\left(
5+2\lambda_{1}^{2}-2\lambda_{2}+\lambda_{2}^{2}+\sqrt{\left(  1+\lambda
_{2}\right)  ^{2}\left(  4\lambda_{1}^{2}+\left(  \lambda_{2}-3\right)
^{2}\right)  }\right)  ~,\label{w4}%
\end{equation}%
\begin{equation}
\bar{\omega}_{2}^{2}=\frac{1}{40}\frac{\Delta_{0}^{2}}{\Delta^{2}}\left(
5+2\lambda_{1}^{2}-2\lambda_{2}+\lambda_{2}^{2}-\sqrt{\left(  1+\lambda
_{2}\right)  ^{2}\left(  4\lambda_{1}^{2}+\left(  \lambda_{2}-3\right)
^{2}\right)  }\right)  ~,\label{w5}%
\end{equation}%
\[
\bar{\omega}_{3}^{2}=\frac{\Delta_{0}^{2}}{\Delta^{2}}\left(  \frac{1}%
{5}+\frac{\lambda_{1}^{2}}{10}\frac{2+\lambda_{2}}{3+\lambda_{2}}\right)  ~,
\]%
\[
\bar{\omega}_{4}^{2}=\frac{1}{20}\frac{\Delta_{0}^{2}}{\Delta^{2}}\left(
\lambda_{1}^{2}+2\lambda_{2}+2\lambda_{2}^{2}+4\right)  ~,
\]%
\[
\bar{\omega}_{5}^{2}=\frac{1}{20}\frac{\Delta_{0}^{2}}{\Delta^{2}}\left(
\left(  1-\lambda_{2}\right)  ^{2}+\frac{\lambda_{1}^{2}}{2\lambda_{2}}\left(
1+3\lambda_{2}\right)  \right)  ~.
\]

By taking into account that $\mathbf{B}_{m_{j}}^{-}=\mathbf{0}$, we have
$\mathbf{B}_{m_{j}}^{\left(  1\right)  }=\left(  -\right)  ^{m_{j}}%
\mathbf{B}_{-m_{j}}^{\left(  2\right)  }$ and%
\begin{equation}
\mathbf{\hat{T}}^{\left(  1\right)  }\left(  \mathbf{n},\omega\right)
=\frac{1}{1+2g_{0}\mathcal{I}_{\mathsf{av}}}\sum_{m_{j}}\mathbf{B}_{m_{j}%
}\left(  \omega\right)  \left(  \hat{\bm{\sigma}}\mathbf{b}_{m_{j}}\right)
\hat{g}~, \label{T1B}%
\end{equation}%
\begin{equation}
\mathbf{\hat{T}}^{\left(  2\right)  }\left(  \mathbf{n},\omega\right)
=\frac{1}{1+2g_{0}\mathcal{I}_{\mathsf{av}}}\sum_{m_{j}}\mathbf{B}_{m_{j}%
}\left(  \omega\right)  \hat{g}\left(  \hat{\bm{\sigma}}\mathbf{b}_{m_{j}%
}\right)  ~. \label{T2B}%
\end{equation}
In the last equation, the property (\ref{b012}) is used.

The full ordinary axial-vector vertices can also be simplified because
$\phi_{l,m}^{-}=0$ and, therefore, $\phi(-\mathbf{n})=\phi(\mathbf{n})$. From
Eqs. (\ref{fipY}), (\ref{fip00}), and (\ref{fi00p}), we find $\phi\left(
\mathbf{n};\omega,\mathbf{q}\right)  =\phi_{0,0}^{+}\left(  \omega
,\mathbf{q}\right)  $ and, thus,
\[
\mathbf{\hat{\tau}}(\mathbf{n},\omega)=\frac{1}{1+2\mathfrak{g}_{0}%
\mathcal{I}_{\mathsf{av}}}\hat{\bm{\sigma}}~,~\mathbf{\hat{\tau}}^{-}\left(
\mathbf{n},\omega\right)  =\frac{1}{1+2\mathfrak{g}_{0}\mathcal{I}%
_{\mathsf{av}}}\hat{\bm{\sigma}}^{T}\mathbf{~.}%
\]

According to modern theories \cite{Khodel}-\cite{0203046}, there are several multicomponent
states that compete in energy depending on the temperature. Accordingly
the phase transitions occur between these states when the temperature goes
down. The possible phase states of the $^{3}P_{2}-^{3}F_{2}$ condensate are
cataloged in Ref. \cite{Clark}. Immediately below the critical temperature, the superfluid condensate can appear either in one of the two two-component phases,
\begin{equation}
O_{\pm3}:~\frac{\Delta_{0}}{\Delta}=\frac{1}{2}\ ,~\lambda_{1}=0~,~\lambda
_{2}=\pm3\label{o3}%
\end{equation}
or in the one-component phase,
\begin{equation}
m_{j}=0:~\Delta_{0}=1,~\lambda_{1}=0,~\lambda_{2}=0~.\label{m0}%
\end{equation}
These lowest-energy states are nearly degenerate. The higher group is composed
of the phases
\begin{equation}
O_{1}:~\frac{\Delta_{0}}{\Delta}=\frac{5}{\sqrt{14}\sqrt{17-3\sqrt{21}}%
},~\lambda_{1}=\frac{3}{5}\sqrt{2\left(  17-3\sqrt{21}\right)  },~\lambda_{2}=\frac{3}{5}\left(  \sqrt{21}-4\right)  \label{o1}%
\end{equation}
and
\begin{equation}
O_{2}:~\frac{\Delta_{0}}{\Delta}=\frac{5}{\sqrt{14}\sqrt{17+3\sqrt{21}}%
},~\lambda_{1}=\frac{3}{5}\sqrt{2\left(  17+3\sqrt{21}\right)  }%
,~\lambda_{2}=-\frac{3}{5}\left(  \sqrt{21}+4\right)  ~.\label{o2}%
\end{equation}
According to the above calculations the energy split between the two groups shrinks
along with the temperature decrease and results in the phase transition at
$T=0.7T_{c}$.

The effective vertices and the polarization tensors for each of the above
phases can be obtained with the aid of Eqs. (\ref{B0})--(\ref{B2}) and
(\ref{PA}). We found that for all of the above-mentioned phases, the anomalous
vertices can be written universally in the form
\begin{equation}
\mathbf{B}_{m_{j}}=\left(  \frac{\Delta_{0}}{\Delta}\right)  ^{2}\frac
{\bar{\omega}}{\bar{\omega}^{2}-1/20}i\left\langle \mathbf{b}_{m_{j}}^{\ast
}\mathbf{\times\bar{b}}\right\rangle \label{B}%
\end{equation}
where%
\begin{align}
\left\langle \mathbf{b}_{0}^{\ast}\mathbf{\times\bar{b}}\right\rangle  &
=-\frac{1}{2\sqrt{6}}\frac{\Delta_{0}}{\Delta}\left(  0,\sqrt{6}\lambda
_{1},0\right)  \,,\nonumber\\
\left\langle \mathbf{b}_{\pm1}^{\ast}\mathbf{\times\bar{b}}\right\rangle  &
=-\frac{1}{2\sqrt{6}}\frac{\Delta_{0}}{\Delta}\left(  -i\left(  3+\lambda
_{2}\right)  ,\mp\left(  3-\lambda_{2}\right)  ,-i\lambda_{1}\right)
~,\nonumber\\
\left\langle \mathbf{b}_{\pm2}^{\ast}\mathbf{\times\bar{b}}\right\rangle  &
=-\frac{1}{2\sqrt{6}}\frac{\Delta_{0}}{\Delta}\left(  \mp i\lambda
_{1},-\lambda_{1},\mp2i\lambda_{2}\right)  ~. \label{bcross}%
\end{align}

Poles of the anomalous vertex indicate eigenmodes of the order parameter. The
pole at $\bar{\omega}^{2}=1/20$ signals the existence of undamped collective
spin oscillations of the energy
\begin{equation}
\omega_{s}\left(  q=0\right)  =\frac{\Delta}{\sqrt{5}}~. \label{ws}%
\end{equation}
Three important conclusions follow immediately from this simple formula: (a)
The spin-density waves are of the identical excitation energy in all
phases of the superfluid neutron liquid. (b) Fermi-liquid interactions do not
influence the spin-wave energy in the condensate. (c) The spin waves are
kinematically able to decay into neutrino pairs through neutral weak currents.

\section{Polarization functions.}

Unfortunately, the Landau parameter $\mathfrak{g}_{0}$ for the particle-hole
interactions in asymmetric nuclear matter is unknown. Therefore, in evaluating
neutrino energy losses, we simply neglect the Fermi-liquid effects by taking
$\mathfrak{g}_{0}\rightarrow0$ and consider the axial polarization in the BCS
approximation. The latter can be obtained in the form \cite{L10a}
\begin{align}
\Pi_{\mathrm{A}}^{ij}\left(  \omega\right)   &  =4\varrho\left\langle \frac
{1}{2}\left(  \mathcal{I}_{GG}-\bar{b}^{2}\mathcal{I}_{FF}\right)  \delta
_{ij}+\bar{b}^{2}\mathcal{I}_{FF}\frac{\bar{b}_{i}\bar{b}_{j}}{\bar{b}^{2}%
}\right\rangle \nonumber\\
&  -\varrho\left\langle \frac{\omega}{2\Delta}\mathcal{I}_{FF}\mathrm{Tr}%
\left[  \hat{\sigma}_{i}\hat{T}_{j}^{\left(  1\right)  }\hat{g}\left(
\hat{\bm{\sigma}}\mathbf{\bar{b}}\right)  -~\hat{\sigma}_{i}\left(
\hat{\bm{\sigma}}\mathbf{\bar{b}}\right)  \hat{g}\hat{T}_{j}^{\left(
2\right)  }\right]  \right\rangle ~, \label{KA}%
\end{align}
where $i,j=1,2,3$ and the anomalous axial-vector vertices $\mathbf{\hat{T}%
}^{\left(  1,2\right)  }$ are given by Eqs. (\ref{T1B}) and (\ref{T2B}). As in
the above, we focus on the case $\mathbf{q=0}$ and omit for brevity the dependence
on $\mathbf{n}$ and $\omega$. By using Eq. (\ref{GGmFF}) and applying a little algebra, we obtain
\begin{align*}
\Pi_{\mathrm{A}}^{ij}\left(  \omega\right)   &  =-4\varrho\left\langle \left(
\delta^{ij}\bar{b}^{2}-\bar{b}^{i}\bar{b}^{j}\right)  \mathcal{I}\right\rangle
\\
&  +\frac{4\varrho\omega^{2}}{\left(  \omega+i0\right)  ^{2}-\Delta^{2}%
/5}\frac{\Delta_{0}^{2}}{\Delta^{2}}\sum_{m_{j}}\left\langle \left(
\mathbf{b}_{m_{j}}\mathbf{\times\bar{b}}\right)  ^{i}\mathcal{I}\right\rangle
\left\langle \mathbf{b}_{m_{j}}^{\ast}\mathbf{\times\bar{b}}\right\rangle
^{j}~.
\end{align*}
The pole location on the complex $\omega$ plane is chosen to obtain the
retarded polarization tensor.

For further evaluation of the polarization function we will again use the
angle-average approximation by replacing the angle-dependent gap in the
quasiparticle energy (\ref{Ep}) with its angle average, $\bar{b}^{2}\Delta
^{2}\rightarrow\left\langle \bar{b}^{2}\Delta^{2}\right\rangle =\Delta^{2}$.
Then the function
\begin{equation}
\mathcal{I}_{\mathsf{av}}\left(  \omega\right)  =\int_{0}^{\infty}%
\frac{d\varepsilon}{E}\frac{\Delta^{2}}{4E^{2}-\left(  \omega+i0\right)  ^{2}%
}\tanh\frac{E}{2T}~ \label{Iav}%
\end{equation}
with $E^{2}=\varepsilon^{2}+4\Delta^{2}$, is isotropic and can be moved beyond
the angle integral. In this approximation, we obtain
\begin{align}
\Pi_{\mathrm{A}}^{ij}\left(  \omega\right)   &  =-4\varrho\left(  \delta
^{ij}-\left\langle \bar{b}^{i}\bar{b}^{j}\right\rangle \right)  \mathcal{I}%
_{\mathsf{av}}\left(  \omega\right) \nonumber\\
&  +\frac{4\varrho\omega^{2}}{\left(  \omega+i0\right)  ^{2}-\Delta^{2}%
/5}\mathcal{I}_{\mathsf{av}}\left(  \omega\right)  \frac{\Delta_{0}^{2}%
}{\Delta^{2}}\sum_{m_{j}}\left\langle \mathbf{b}_{m_{j}}\mathbf{\times\bar{b}%
}\right\rangle ^{i}\left\langle \mathbf{b}_{m_{j}}^{\ast}\mathbf{\times\bar
{b}}\right\rangle ^{j}~. \label{PA}%
\end{align}
Below we use the retarded polarization tensor for calculation of the neutrino
energy losses from superfluid bulk matter of neutron stars. In this
calculation one can neglect the temporal and mixed components of the tensor
occurring as small relativistic corrections \cite{L09}.

\section{Neutrino energy losses$\allowbreak\allowbreak$}

We will examine the neutrino energy losses in the standard model of weak
interactions. Then after integration over the phase volume of freely escaping
neutrinos and antineutrinos the total energy which is emitted per unit volume
and time can be obtained in the form (see details, e.g., in Ref. \cite{L01})
\begin{equation}
\epsilon=-\frac{G_{F}^{2}C_{A}^{2}\mathcal{N}_{\nu}}{192\pi^{5}}\int
_{0}^{\infty}d\omega\int d^{3}q\frac{\omega\Theta\left(  \omega-q\right)
}{\exp\left(  \frac{\omega}{T}\right)  -1}\operatorname{Im}\Pi_{\mathrm{A}%
}^{\mu\nu}\left(  \omega,\mathbf{q}\right)  \left(  k_{\mu}k_{\nu}-k^{2}%
g_{\mu\nu}\right)  ~, \label{QQQ}%
\end{equation}
where $G_{F}$ is the Fermi coupling constant, $C_{A}=1.26$ is the axial-vector
weak coupling constant of neutrons, $\mathcal{N}_{\nu}=3$ is the number of
neutrino flavors, $\Theta\left(  x\right)  $ is the Heaviside step function,
and $k^{\mu}=\left(  \omega,\mathbf{q}\right)  $ is the total energy and
momentum of the freely escaping neutrino pair $\left(\mu,\nu=0,1,2,3\right)$.

In Eq. (\ref{QQQ}), we have neglected the neutrino emission in the vector
channel, which is strongly suppressed due to conservation of the vector
current. Therefore the energy losses are connected to the imaginary part of
the retarded polarization tensor in the axial channel, $\Pi_{\mathrm{A}}%
^{\mu\nu}\simeq\delta^{\mu i}\delta^{\nu j}\operatorname{Im}\Pi_{\mathrm{A}%
}^{ij}$. The latter is caused by the pair breaking and formation (PBF)
processes and by the spin-wave decays (SWDs). These processes operate in
different kinematical domains, so that the imaginary part of the polarization
tensor consists of two clearly distinguishable contributions, $\operatorname{Im}%
\Pi_{\mathrm{A}}^{ij}=\operatorname{Im}\Pi_{\mathrm{PBF}}^{ij}%
+\operatorname{Im}\Pi_{\mathrm{SWD}}^{ij}$, which we will now consider.

\subsection{PBF channel}

The imaginary part of $\mathcal{I}_{\mathsf{av}}$, which arises from the poles
of the integrand in Eq. (\ref{Iav}) at $\omega=\pm2E$ is given by%
\[
\operatorname{Im}\mathcal{I}_{\mathsf{av}}\left(  \omega>0\right)  =\frac{\pi
}{2}\frac{\Delta^{2}\Theta\left(  \omega^{2}-4\Delta^{2}\right)  }{\omega
\sqrt{\omega^{2}-4\Delta^{2}}}\tanh\frac{\omega}{4T}~.
\]
The PBF processes occur if $\omega>2\Delta$, while the energy (\ref{ws}) of the
spin waves is considerably smaller. Therefore we may neglect the term
$\allowbreak\Delta^{2}/5$ relative to $\omega^{2}$ in the denominator of Eq.
(\ref{PA}). Then,
\begin{align}
\operatorname{Im}\Pi_{ij}^{\mathrm{PBF}}\left(  \omega\right)   &
=-2\pi\varrho\left(  \delta_{ij}-\left\langle \bar{b}_{i}\bar{b}%
_{j}\right\rangle -\frac{\Delta_{0}^{2}}{\Delta^{2}}\Lambda_{ij}\right)
\nonumber\\
&  \times\frac{\Delta^{2}\Theta\left(  \omega^{2}-4\Delta^{2}\right)  }%
{\omega\sqrt{\omega^{2}-4\Delta^{2}}}\tanh\frac{\omega}{4T}~, \label{IPA}%
\end{align}
where
\[
\Lambda_{ij}\equiv\sum_{m_{j}}\left\langle \mathbf{b}_{m_{j}}\mathbf{\times
\bar{b}}\right\rangle _{i}\left\langle \mathbf{b}_{m_{j}}^{\ast}%
\mathbf{\times\bar{b}}\right\rangle _{j}%
\]
The tensors $\left\langle \bar{b}_{i}\bar{b}_{j}\right\rangle $ and
$\Lambda_{ij}$ can be found with the aid of Eqs. (\ref{burb}) and
(\ref{bcross}). A straightforward calculation gives%
\begin{equation}
\left\langle \bar{b}_{i}\bar{b}_{i}\right\rangle =\frac{1}{6}\frac{\Delta
_{0}^{2}}{\Delta^{2}}\left(
\begin{array}
[c]{ccc}%
1-2\lambda_{2}+\lambda_{2}^{2}+\lambda_{1}^{2} & 0 & -\lambda_{1}-\lambda
_{1}\lambda_{2}\\
0 & 1+2\lambda_{2}+\lambda_{2}^{2} & 0\\
-\lambda_{1}-\lambda_{1}\lambda_{2} & 0 & 4+\lambda_{1}^{2}%
\end{array}
\right)  ~, \label{bb}%
\end{equation}
and%
\begin{equation}
\Lambda_{ij}=\frac{1}{12}\frac{\Delta_{0}^{2}}{\Delta^{2}}\left(
\begin{array}
[c]{ccc}%
\lambda_{1}^{2}+\left(  \lambda_{2}+3\right)  ^{2} & 0 & 3\lambda_{1}\left(
\lambda_{2}+1\right) \\
0 & 4\lambda_{1}^{2}+\left(  \lambda_{2}-3\right)  ^{2} & 0\\
3\lambda_{1}\left(  \lambda_{2}+1\right)  & 0 & \lambda_{1}^{2}+4\lambda
_{2}^{2}%
\end{array}
\right)  ~. \label{LAM}%
\end{equation}

Inserting the imaginary part of the polarization tensor into Eq. (\ref{QQQ}),
we calculate the contraction of $\operatorname{Im}\Pi_{\mathrm{PBF}}^{\mu\nu}$
with the symmetric tensor $k_{\mu}k_{\nu}-k^{2}g_{\mu\nu}$. This gives%
\begin{align}
\epsilon &  =\frac{1}{96\pi^{6}}G_{F}^{2}C_{A}^{2}\mathcal{N}_{\nu}%
p_{F}M^{\ast}\Delta^{2}\int_{0}^{\infty}\frac{d\omega}{\left(  1+\exp
\frac{\omega}{2T}\right)  ^{2}}\frac{\Theta\left(  \omega^{2}-4\Delta
^{2}\right)  }{\sqrt{\omega^{2}-4\Delta^{2}}}\nonumber\\
&  \times\int\limits_{q<\omega}d^{3}q\left(  2\omega^{2}-q^{2}-q_{i}%
\left\langle \bar{b}_{i}\bar{b}_{i}\right\rangle q_{j}-\frac{3}{2}\frac
{\Delta_{0}^{2}}{\Delta^{2}}\left(  \omega^{2}-q^{2}\right)  -\frac{\Delta
_{0}^{2}}{\Delta^{2}}q_{i}\Lambda_{ij}q_{j}\right)  ~.\label{Enu}%
\end{align}
Integration over $d^{3}q$ can be done in cylindrical frame, where
$q_{1}=q_{\perp}\cos\Phi$, $q_{2}=q_{\perp}\sin\Phi$, and $q_{3}=q_{z}$. This results
in the neutrino energy losses in the form%
\begin{equation}
\epsilon=\frac{1}{60\pi^{5}}\left(  1-\frac{3}{4}\frac{\Delta_{0}^{2}}%
{\Delta^{2}}\right)  G_{F}^{2}C_{A}^{2}\mathcal{N}_{\nu}p_{F}M^{\ast}%
\int_{2\Delta}^{\infty}\frac{\omega^{5}d\omega}{\left(  1+\exp\frac{\omega
}{2T}\right)  ^{2}}\frac{\Delta^{2}}{\sqrt{\omega^{2}-4\Delta^{2}}%
}~~.\label{Enu1}%
\end{equation}
In obtaining this expression, the following fact is used:
\[
\frac{2}{3}\operatorname*{Tr}\Lambda_{ij}=\operatorname*{Tr}\left\langle
\bar{b}_{i}\bar{b}_{i}\right\rangle =\frac{\Delta_{0}^{2}}{\Delta^{2}}\left(
\frac{1}{3}\lambda_{1}^{2}+\frac{1}{3}\lambda_{2}^{2}+1\right)  \equiv1~.
\]
With the aid of the change $\omega=2T\sqrt{x^{2}+\Delta^{2}/T^{2}}$, one can
recast Eq. (\ref{Enu1}) into the form:%
\begin{equation}
\epsilon_{\mathrm{PBF}}=\frac{2}{15\pi^{5}}\left(  4-3\frac{\Delta_{0}^{2}%
}{\Delta^{2}}\right)  G_{F}^{2}C_{A}^{2}\mathcal{N}_{\nu}p_{F}M^{\ast}%
T^{7}y^{2}\int_{0}^{\infty}\frac{z^{4}dx}{\left(  1+\exp z\right)  ^{2}%
}\label{PBF}%
\end{equation}
where $z=\sqrt{x^{2}+y^{2}}$ and $y=\Delta\left(  T\right)  /T$.

For a practical usage from Eq. (\ref{PBF}), we find
\begin{equation}
\epsilon_{\mathrm{PBF}}=5.\,\allowbreak85\times10^{20}~\left(  \frac{M^{\ast}%
}{M}\right)  \left(  \frac{p_{F}}{Mc}\right)  T_{9}^{7}\mathcal{N}_{\nu
}C_{\mathrm{A}}^{2}F_{\mathrm{PBF}}\left(  y\right)  ~~~\frac{erg}{cm^{3}%
s}~,\label{ergPBF}%
\end{equation}
where $M$ and $M^{\ast}=p_{F}/\upsilon_{F}$ are the bare and effective
nucleon masses, respectively; $T_{9}=T/10^{9}\mathrm{K}$, and
\begin{equation}
F_{\mathrm{PBF}}\left(  y\right)  =\left(  4-3\frac{\Delta_{0}^{2}}{\Delta
^{2}}\right)  y^{2}\int_{0}^{\infty}dx\frac{z^{4}}{\left(  1+\exp z\right)
^{2}}\text{.}\label{Ft}%
\end{equation}
The neutrino energy losses, as given by Eq. (\ref{PBF}) with good accuracy
reproduce the result obtained in Ref. \cite{L10a} for the one-component
phase $m_{j}=0$, where $\Delta_{0}/\Delta=1$. It is necessary to notice that
Eq. (\ref{Ft}) obtained in the angle-average approximation is much simpler for
numerical evaluation than the "exact" expression which contains additionally
the angle integration \cite{L10a}. To avoid possible misunderstanding we
stress that the gap amplitude $\Delta\left(  T\right)  $ in Eq. (\ref{Ft}) is
$\sqrt{2}$ times larger than the gap amplitude $\Delta_{YKL}$ used in Ref.
\cite{YKL} , where the same anisotropic gap $\Delta_{\mathbf{n}}=\Delta\bar
{b}\left(  \mathbf{n}\right)  $ is written in the form $\Delta_{\mathbf{n}%
}=\Delta_{YKL}\sqrt{1+3\cos^{2}\theta}\equiv\Delta_{YKL}\sqrt{2}\,\bar
{b}\left(  \mathbf{n}\right)  $. In other words, $\left\langle \Delta
_{\mathbf{n}}^{2}\right\rangle =\Delta^{2}=2\Delta_{YKL}^{2}$.

The small difference in the gap amplitudes, $\delta\Delta^{2}/\Delta^{2}%
\sim2\%$, inherent for various phases of the condensate is crucial for the
phase transitions \cite{0203046}, but this small inequality can be disregarded
in evaluation of the neutrino energy losses. Therefore the efficiency of PBF
processes in various phases of the superfluid condensate is proportional to
the factor%
\[
\kappa_{\mathrm{PBF}}\equiv1-\frac{3}{4}\frac{\Delta_{0}^{2}}{\Delta^{2}}%
\]
For the one-component phase $m_{j}=0$ one has $\kappa_{\mathrm{PBF}}=1/4$; in
the case of $O_{\pm3}$ condensates, $\kappa_{\mathrm{PBF}}=13/16$. For the
$O_{1}$ phase, we obtain $\kappa_{\mathrm{PBF}}=\left(  173-9\sqrt{21}\right)
/224\simeq 0.5882$, and for the $O_{2}$ phase, we have $\kappa
_{\mathrm{PBF}}=\left(  173+9\sqrt{21}\right)  /224\simeq 0.95644$.

\subsection{SWD channel}

In the frequency domain $0<\omega<\Delta\bar{b}$, the imaginary part of the
weak polarization tensor (\ref{PA}) arises from the pole of the denominator at
$\omega=\Delta/\sqrt{5}$ and can be written as%
\begin{equation}
\operatorname{Im}\Pi_{ij}^{\mathrm{SWD}}\left(  \omega>0\right)  =-\frac{4}%
{5}\pi\sqrt{5}\Delta\varrho\delta\left(  \omega-\Delta/\sqrt{5}\right)
\mathcal{I}_{\mathsf{av}}\left(  \omega\right)  \frac{\Delta_{0}^{2}}%
{\Delta^{2}}\Lambda_{ij}~. \label{ISWD}%
\end{equation}
$\allowbreak\allowbreak$Inserting this expression into Eq. (\ref{QQQ}) and
performing trivial calculations, we find%
\begin{equation}
\epsilon_{\mathrm{SWD}}=\frac{G_{F}^{2}C_{A}^{2}\mathcal{N}_{\nu}}{160\pi^{5}%
}p_{F}M^{\ast}\frac{\Delta_{0}^{2}}{\Delta^{2}}\frac{\left(  \Delta/\sqrt
{5}\right)  ^{7}}{\exp\left(  \frac{\Delta/\sqrt{5}}{T}\right)  -1}\int
_{0}^{\infty}\frac{d\varepsilon}{E}\frac{\Delta^{2}}{E^{2}-\Delta^{2}/20}%
\tanh\frac{E}{2T}~.\label{Eswd}%
\end{equation}
One can neglect $\Delta^{2}/20$ in comparison to $E^{2}>4\Delta^{2}$ in the
denominator of the integrand. With this simplification, we obtain%
\begin{equation}
\epsilon_{\mathrm{SWD}}=\frac{G_{F}^{2}\mathcal{N}_{\nu}}{160\pi^{5}}C_{A}%
^{2}p_{F}M^{\ast}T^{7}\frac{1}{\exp\left(  y/\sqrt{5}\right)  -1}\left(
\frac{y}{\sqrt{5}}\right)  ^{7}\mathcal{I}_{0}\left(  y\right)  ~,\label{SWD}%
\end{equation}
where $y\equiv\Delta\left(  T\right)  /T$, and%
\begin{equation}
\mathcal{I}_{0}\left(  y\right)  =\int_{0}^{\infty}\frac{du}{\left(
u^{2}+1\right)  ^{3/2}}\tanh\frac{y}{2}\sqrt{u^{2}+1}~.\label{I0}%
\end{equation}
This expression can be recast into the traditional form
\begin{equation}
\epsilon_{\mathrm{SWD}}=2.\,\allowbreak74\times10^{19}~\left(  \frac{M^{\ast}%
}{M}\right)  \left(  \frac{p_{F}}{Mc}\right)  T_{9}^{7}\mathcal{N}_{\nu
}C_{\mathrm{A}}^{2}\frac{\Delta_{0}^{2}}{\Delta^{2}}\frac{\left(  y/\sqrt
{5}\right)  ^{7}\mathcal{I}_{0}\left(  y\right)  }{\exp\left(  y/\sqrt
{5}\right)  -1}~~~\frac{erg}{cm^{3}s}~.\label{ESWD}%
\end{equation}
From this expression, it is seen that the efficieency of spin-wave decays is
proportional to $\kappa_{\mathrm{SWD}}=\Delta_{0}^{2}/\Delta^{2}$. For the
one-component condensate $m_{j}=0$, $\kappa_{\mathrm{SWD}}=1$; in the
case of $O_{\pm3}$ condensates, $\kappa_{\mathrm{SWD}}=1/4$. For the $O_{1}$
phase, we obtain $\kappa_{\mathrm{SWD}}=\left(  17+3\sqrt{21}\right)
/56\simeq\allowbreak0.549\,07$, and for the $O_{2}$ phase, 
$\kappa_{\mathrm{SWD}}=\left(  17-3\sqrt{21}\right)  /56\simeq\allowbreak
5.81\times10^{-2}$.

\section{Phase transitions and efficiency of neutrino emission}

For numerical evaluation of the neutrino losses, it is necessary to know the
function $y=\Delta\left(  T\right)  /T$, which in general is to be found with
the aid of gap equations. However, as mentioned above, the difference in the
gap amplitudes for various phases can be neglected in evaluation of the
neutrino energy losses. This substantially simplifies the problem because for
the case $m_{j}=0$ the function is well investigated. We can adjust, for
example, the simple fit to $\Delta_{YKL}\left(  T\right)  /T=\mathsf{v}%
_{B}\left(  \tau\right)  $, as suggested in Ref. \cite{YKL}, where $\tau\equiv
T/T_{c}$. Taking into account that, in Ref. \cite{YKL}, the gap amplitude
$\Delta_{YKL}\left(  T\right)  $ is defined by the relation $\Delta
_{\mathbf{n}}^{2}=\Delta_{YKL}^{2}\left(  1+3\cos^{2}\theta\right)  $, while
our definition is $\Delta_{\mathbf{n}}^{2}=\frac{1}{2}\Delta^{2}\left(
1+3\cos^{2}\theta\right)  $, we obtain $y\left(  \tau\right)  =\sqrt
{2}\mathsf{v}_{B}\left(  \tau\right)  $.

\begin{figure}[ptb]
\includegraphics{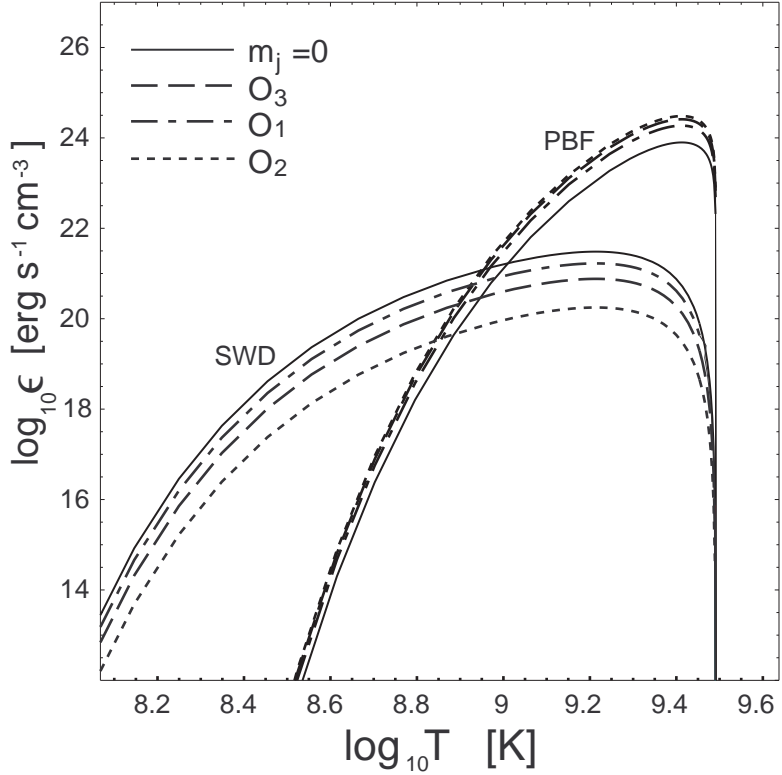}\caption{Temperature dependence of the neutrino
emissivity due to recombination of Cooper pairs (PBF) and due to decay of spin
waves (SWD) for $p_{F}=2.1~fm^{-1}$ and $T_{c}=3\times10^{9}$ K. The effective
mass is taken to be $M^{\ast}=0.7M$. The curves of different styles correspond
to various phases of superfluid condensate, which are discussed in the text.}%
\label{fig2}%
\end{figure}

In Fig. \ref{fig2} we compare the PBF and SWD neutrino emissivity for various
phases of superfluid neutron matter. The temperature dependence of the
emissivity is evaluated at $p_{F}=2.1fm^{-1}$. We set the effective nucleon
masses $M^{\ast}=0.7M$; the critical temperature for neutron pairing is chosen
to be $T_{c}=3\times10^{9}K$.

One can see that the decay of spin waves into neutrino pairs is very effective
at low temperatures, when other known mechanisms of neutrino energy losses in
the bulk neutron matter are strongly suppressed by superfluidity. Maximal
neutrino emission in the SWD channel occurs in the one-component phase. In
contrast, efficiency of the PBF channel is maximal in the $O_{2}$ phase.

\begin{figure}[ptb]
\includegraphics{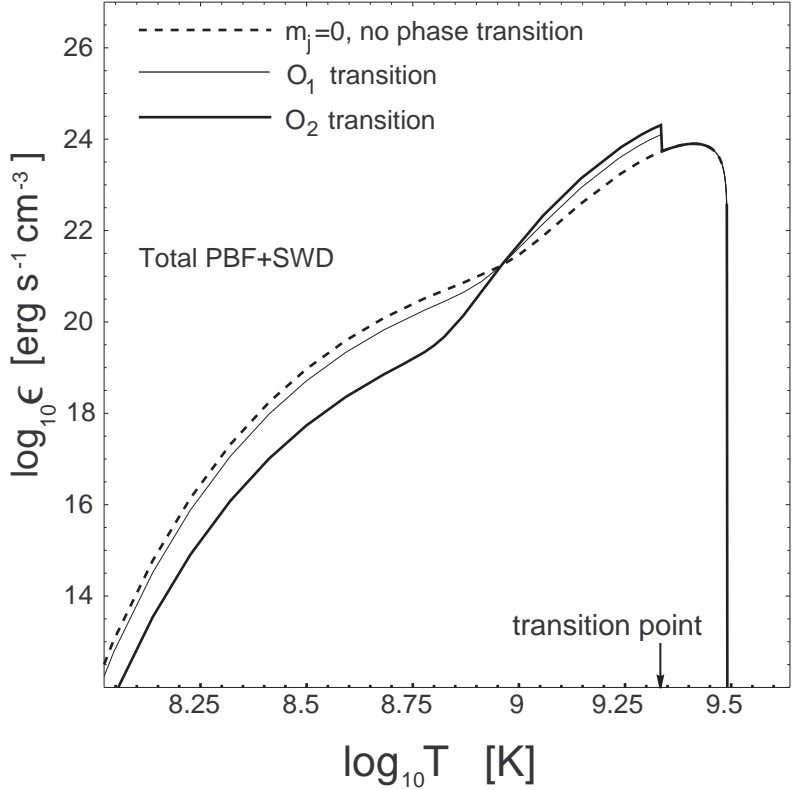}\caption{Total energy losses $\epsilon
=\epsilon_{\mathrm{PBF}}+\epsilon_{\mathrm{SWD}}$ versus the temperature. The
superfluid condensate undergoes the phase transition at $T=0.7T_{c}$.
Parameters of the medium are the same as in Fig. 2.}%
\label{fig3}%
\end{figure}

In Fig. \ref{fig3} we demonstrate the total neutrino emissivity $\epsilon
=\epsilon_{\mathrm{PBF}}+\epsilon_{\mathrm{SWD}}$ versus the temperature by
assuming that the phase transition occurs at $T=0.7T_{c}$. The phase
transition (if it occurs) leads to a sharp increase in the neutrino energy
losses followed by a decrease, along with a decrease in the temperature that takes place more rapidly than it would without the phase transition.

According to the minimal cooling paradigm \cite{Page04}, \cite{Page09}, along with lowering of the temperature, the star continues to lose its energy by radiating low-energy neutrinos via the PBF processes untill a photon-cooling epoch enters at 
$T\sim 0.1T_{c}$. At this latest stage of the cooling, all mechanisms of neutrino emission from the inner core are suppressed greatly by the neutron and proton superfluidity, and the $\gamma$ radiation from the star surface is considered as the main mechanism of the star cooling. 

\begin{figure}[ptb]
\includegraphics{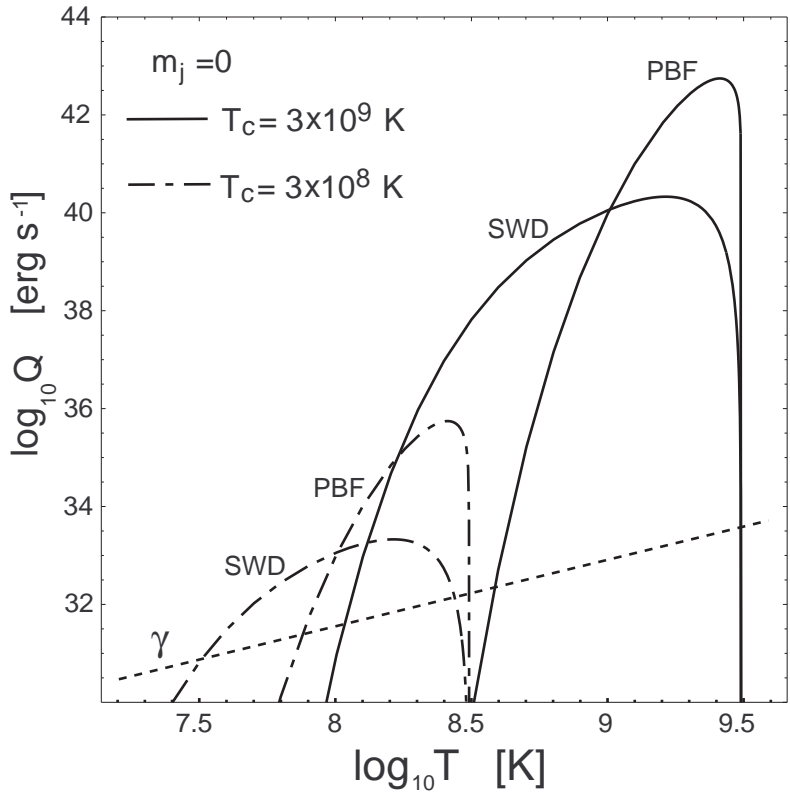}\caption{Temperature dependence of the neutrino
luminosities from the model neutron star due to recombination of Cooper pairs (PBF) and due to decay of spin waves (SWD) for $p_{F}=2.1~fm^{-1}$. 
The effective mass is taken to be $M^{\ast}=0.7M$. Solid lines correspond to $T_{c}=3\times10^{9}$ K; dot-dashed lines correspond to $T_{c}=3\times10^{8}$ K. Volume of the triplet condensate is estimated as $7\times 10^{18}\ cm^3$.  The dashed line is the energy losses per unit of time due to surface $\gamma$ radiation, as calculated in Ref. \cite{Page04}.}%
\label{fig4}%
\end{figure}

Given the strong dependence of the PBF and SWD neutrino emission on the temperature and the density, the overall effect of the SWD processes can only be assessed by complete calculations of the neutron star cooling which are beyond the scope of this paper.  A rough  estimate can be made by considering a simplified model of the superfluid core of the density $2\rho_0$ enclosed in the volume $7\times 10^{18}\ cm^3$. In Fig. \ref{fig4} we demonstrate the total neutrino energy losses caused by PBF and SWD neutrino emission from the star volume in comparison with the surface photon radiation. The latter is taken as in Fig. 20 of Ref. \cite{Page04}. There is a relatively large range of predicted values for $T_{c}$; therefore, we show the PBF and SWD neutrino luminosities for $T_{c}=3\times10^{9}$ K and for $T_{c}=3\times10^{8}$ K. This simple estimate shows that the neutrino emission caused by spin-wave decay (SWD) can dominate the $\gamma$ radiation within some temperature range, which was previously considered as the photon-cooling era.

\begin{figure}[ptb]
\includegraphics{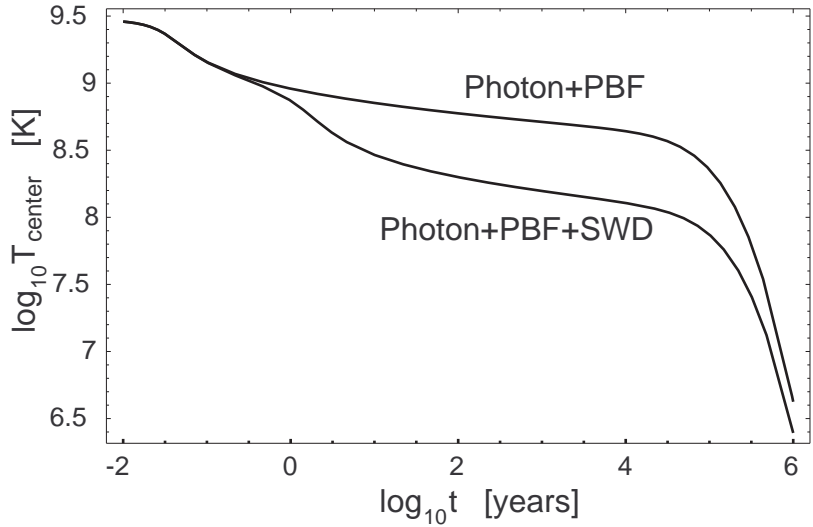}\caption{Transition of cooling trajectories between a model with a surface photon radiation and PBF neutrino emission and a model including a surface photon radiation and PBF+SWD neutrino processes. 
$T_{c}=3\times10^{9}$ K. Details of the calculation are described in the text.}%
\label{fig5}%
\end{figure}
 
To get an idea of the lowering of the cooling trajectory due to SWD neutrino emission we consider a simple model of cooling of the above core of the density $2\rho_0$ and of the volume $7\times 10^{18}\ cm^3$ enclosed in a thin envelope typical for real neutron stars by assuming that the surface temperature is connected to the central temperature by the formula in Ref. \cite{Gudm}. We assume also that the bulk matter consists mostly of $^3P_2$ superfluid neutrons with $m_j=0$ and contains a small admixture of normal (nonsuperfluid) protons and electrons (the proton fraction $x_{p}=0.05$), so that the total specific heat $C_{v}$ consists of the three corresponding contributions, as described in Ref. \cite{Yak99}. Under these conditions, the cooling equation 
\begin{equation}
C_{v}\frac{dT}{dt}=-Q \ ,
\end{equation}
can be solved numerically with $T=T_{c}$ at the initial  moment. We have solved this equation for the case when $Q=Q_{\gamma}+Q_{PBF}$ and for the case when $Q=Q_{\gamma}+Q_{PBF}+Q_{SWD}$. The result is shown in Fig. \ref{fig5}.

\section{Summary and conclusion}

Let us summarize our results. We have studied the linear response of the
superfluid neutron liquid to an external axial-vector field. The calculation is
made for the case of a multicomponent condensate involving several magnetic
quantum numbers and allows us to consider various phases of superfluid neutron
liquid. In order to estimate the neutrino energy losses, while taking into
account possible phase transitions, we have considered the low-energy
excitations of the multicomponent condensate.

Along with the well-known excitations in the form of broken Cooper pairs, we
consider the collective waves of spin density, which are known to exist in the
one-component condensate at very low energy \cite{L10b}. Our theoretical
analysis predicts the existence of such waves in all of the multicomponent
phases we have considered. We found that the excitation energy of spin waves
is identical for all of the phases and is independent of the Fermi-liquid
interactions. In the angle-average approximation, the energy of spin-density
oscillations is estimated as $\omega_{s}\left(  q=0\right)  \simeq\Delta
/\sqrt{5}$.

Neutrino energy losses caused by the pair recombination and spin-wave decays
are given by Eqs. (\ref{PBF}) and (\ref{SWD}), respectively. Because of a
rather small excitation energy, the decay of spin waves leads to a substantial
neutrino emission at the lowest temperatures $T\ll T_{c}$, when all other
mechanisms of the neutrino energy losses are killed by a superfluidity. We
have evaluated the neutrino energy losses for all of the multicomponent phases
that might represent the ground state of the condensate according to modern
theories.

Finally we have evaluated the temperature dependence of neutrino energy losses
from the superfluid neutron liquid in the case when the phase transition
occurs in the condensate at the temperature $T=0.7T_{c}$. Our estimate
predicts a sharp increase of the neutrino energy losses followed by a decrease, along with a decrease of the temperature that takes place more rapidly than it would without the phase transition.

Since the neutron triplet-spin pairing occurs in the core which contains more
than 90\% of the neutron star volume, the neutrino processes discussed here
could influence the evolution of neutron stars.

\end{document}